\magnification=1200
\centerline
{SINGULARITIES OF THE SUSCEPTIBILITY OF AN SRB MEASURE}
\centerline
{IN THE PRESENCE OF STABLE-UNSTABLE TANGENCIES.
\footnote{*}{An early version of this work was presented at the France-Brazil mathematical conference at IMPA in Sep. 2009, at the LAWNP conference in Buzios in Oct. 2009, and at the meeting {\it Progress in Dynamics} at the IHP in Nov. 2009.}}
\bigskip\bigskip
\centerline{by David Ruelle\footnote{$\dagger$}{Math. Dept., Rutgers University, and 
IHES, 91440 Bures sur Yvette, France. email: ruelle@ihes.fr}.}
\bigskip\bigskip\bigskip\bigskip\noindent
	{\leftskip=2cm\rightskip=2cm\sl Abstract. Let $\rho$ be an SRB (or ``physical''), measure for the discrete time evolution given by a map $f$, and let $\rho(A)$ denote the expectation value of a smooth function $A$.  If $f$ depends on a parameter, the derivative $\delta\rho(A)$ of $\rho(A)$ with respect to the parameter is formally given by the value of the so-called susceptibility function $\Psi(z)$ at $z=1$.  When $f$ is a uniformly hyperbolic diffeomorphism, it has been proved that the power series $\Psi(z)$ has a radius of convergence $r(\Psi)>1$, and that $\delta\rho(A)=\Psi(1)$, but it is known that $r(\Psi)<1$ in some other cases.  One reason why $f$ may fail to be uniformly hyperbolic is if there are tangencies between the stable and unstable manifolds for $(f,\rho)$.  The present paper gives a crude, nonrigorous, analysis of this situation in terms of the Hausdorff dimension $d$ of $\rho$ in the stable direction.  We find that the tangencies produce singularities of $\Psi(z)$ for $|z|<1$ if $d<1/2$, but only for $|z|>1$ if $d>1/2$.  In particular, if $d>1/2$ we may hope that $\Psi(1)$ makes sense, and the derivative $\delta\rho(A)=\Psi(1)$ has thus a chance to be defined.\par}
\vfill\eject
\noindent
{\bf 0. Introduction.}
\medskip
	Let $f$ be a diffeomorphism of the compact manifold $M$, and $\rho$ an SRB measure\footnote{$^1$}{For a discussion of SRB measures (Sinai-Ruelle-Bowen) see for instance [10], [27], [2], and references given there.  For recent work analyzing SRB measures for a class of noninvertible maps, see [1].} for $f$.  The derivative $\delta_X\rho(A)$ of the map $f\mapsto\rho$ in the direction of the smooth vector field\footnote{$^2$}{If we replace $f:x\mapsto fx$ by $x\mapsto fx+X(fx)$, then $\rho$ is replaced by $\rho+\delta_X\rho$ to first order in $X$.  The derivative of $f\mapsto\rho$ in the direction of $X$, evaluated at $A$, is $\delta_X\rho(A)$.} ${\bf X}$, evaluated at the smooth real function $A$, can be formally computed to be the value at $z=1$ of
$$	\Psi(z)=\sum_{n=0}^\infty z^n\int\rho(dx)\,{\bf X}(x)\cdot\partial_x(A\circ f^n)
	\eqno{(1)}      $$
We shall call $\Psi$ the {\it susceptibility}\footnote{$^3$}{The physical susceptibility is defined for a continuous time dynamical system, and is a function of the frequency $\omega$.  For the discrete time dynamics considered here, the susceptibility would be $\omega\mapsto\Psi(e^{i\omega})$, but for simplicity we call $\Psi$ the susceptibility.}.
\medskip
	In the uniformly hyperbolic case (i.e., if the support of $\rho$ is a mixing Axiom A attractor for $f$), $\Psi$ has a radius of convergence $r(\Psi)>1$.  One can furthermore prove that $f\mapsto\rho$ is differentiable and that its derivative is given by $\Psi(1)$\footnote{$^4$}{The differentiability of $f\mapsto\rho$ has been established in [9], the inequality $r(\Psi)>1$ and the identity $\delta_X\rho(A)=\Psi(1)$ are proved in [16].  There are corresponding results for hyperbolic flows [18], [3], and generalizations to partially hyperbolic systems [5].}.  In nonuniformly hyperbolic situations these assertions may fail: $r(\Psi)$ may be $<1$, and $f\mapsto\rho$ is presumed to be nondifferentiable\footnote{$^5$}{$r(\Psi)<1$ has been proved for certain (noninvertible) unimodal maps of the interval [17], [8], see also [19] and work in progress by Baladi and Smania.  The analysis in [19] strongly suggests that for a certain class of unimodal maps, the function $f\mapsto\rho$ is nondifferentiable, even in the (weak) Whitney sense.  There is also numerical evidence [4] that $r(\Psi)<1$ for the classical H\'enon attractor.  For recent work on H\'enon-like diffeomorphisms, see [14].}.
\medskip
	The above results suggest two problems: I. proving that $r(\Psi)<1,\ge1,$ or $>1$ in cases of some generality, and II. relating the derivative of $f\mapsto\rho$ to $\Psi(1)$ when this quantity is defined.  The present note is about the first problem, and presents a nonrigorous study of the singularities of $\Psi$ which may occur as a result of {\it tangencies}, i.e., tangencies of stable and unstable manifolds for the system $(f,\rho)$, assumed to have no zero Lyapunov exponent.  (The existence of tangencies excludes uniform hyperbolicity).  Our study is not rigorous, but suggests that $r(\Psi)<1$ if the partial Hausdorff dimension $d$ of $\rho$ in the stable direction is $<{1\over2}$, while $r(\Psi)\ge1$ if $d\ge1/2$.  This opens the possibility that, for some {\it fat} tangencies ($d$ sufficiently large), $\Psi(1)$ is well defined.  In that case, a derivative of $f\mapsto\rho$ may exist, with applications to the physical theory of {\it linear response}\footnote{$^6$}{A basic  physical article on linear response is [21].  A review of linear response for dynamical systems is given in [20].}.
\medskip\noindent
{\bf Acknowledgments.}
\medskip
	I am indebted to Artur \`Avila, Viviane Baladi, Bruno Cessac, Jean-Pierre Eckmann, and Jean-Christophe Yoccoz for valuable remarks.
\medskip\noindent
{\bf 1. Example: volume preserving diffeomorphisms.}
\medskip
	Let $\ell$ be equivalent to Lebesgue measure on $M$, and let the $f$-invariant probability measure $\rho$ be the restriction of $\ell$ to a certain open set $S\subset M$
[similarly, we may also consider the situation where $f$ acts on ${\bf R}^m$ , and $S$ is a bounded open set in ${\bf R}^m$].  If either supp$X\subset S$, or supp$A\subset S$, we may write
$$	\Psi(z)
	=\sum_{n=0}^\infty z^n\int_S\ell(dx)\,{\bf X}(x)\cdot\partial_x(A\circ f^n)
	=-\sum_{n=0}^\infty z^n\int_S\ell(dx)\,[{\rm div}_\ell{\bf X}(x)]A(f^nx)     $$
and therefore $r(\Psi)\ge1$.
\medskip
	If we furthermore suppose that either supp$X\subset S$, or supp$A\subset S$ and $\ell(A)=0$, we have
$$	(f,\rho)\,{\rm exponentially}\,{\rm mixing}\quad
	\Rightarrow\quad r(\Psi)>1  $$
Volume preserving Anosov diffeomorphisms satisfy this condition, and the same is true of the time 1 map of an exponentially mixing volume preserving Anosov flow (which is not uniformly hyperbolic).  Can exponential mixing happen for non-Anosov area preserving diffeomorphisms in 2 dimensions?  We shall now see that mixing already implies that $\Psi(1)$ is well defined, and $\delta_X\rho(A)=\Psi(1)$ when the derivative $\delta_X$ is taken along diffeomorphisms preserving a (parameter dependent) volume.
\medskip
	For simplicity we discuss the case $S=M$.  Let $\rho$ be a probability measure equivalent to Lebesgue measure on the compact manifold $M$.  Denote by $\tilde\rho$ the density of $\rho$ with respect to Lebesgue measure on some charts.  Thus $(f^*\rho)^\sim=\tilde\rho\,\circ f^{-1}/J\circ f^{-1}$ where $J(x)=|{\rm det}(D_xf)|$.  Suppose now that $f,\tilde\rho$ depend smoothly on a parameter, and denote the derivative with respect to the parameter by a {\it prime}.  In particular $f'=X\circ f, J'(x)=J(x)[{\rm div}X(fx)]$.
\medskip
	Writing $\rho_1=f^*\rho$ we have $\tilde\rho_1=(\tilde\rho/J)\circ f^{-1}$, or $\tilde\rho(x)=J(x)(\tilde\rho_1(fx))$, hence
$$	\tilde\rho'(x)=J(x)[\tilde\rho_1'(fx)+\partial_{fx}\tilde\rho_1\cdot X(fx)]
	+J(x)[{\rm div}X(fx)]\tilde\rho_1(fx)  $$
or
$$	(\tilde\rho'/J)\circ f^{-1}
	=\tilde\rho_1'+\partial\tilde\rho_1\cdot X+[{\rm div}X]\tilde\rho_1
	=\tilde\rho_1'+{\rm div}(\tilde\rho'_1X)
	=\tilde\rho_1'+[{\rm div}_{\tilde\rho_1}X]\tilde\rho_1  $$
hence
$$	\int dx\,\tilde\rho'(x)A(fx)=\int dx\,\tilde\rho_1'(x)A(x)
	+\int dx\,\tilde\rho_1(x)[{\rm div}_{\tilde\rho_1}X(x)]A(x)  $$
Imposing the invariance condition $\rho=f^*\rho$, we have thus
$$	\int dx\,\tilde\rho'(x)A(f^{N+1}x)=\int dx\,\tilde\rho'(x)A(x)
	+\sum_{n=0}^N\int dx\,\tilde\rho(x)[{\rm div}_{\tilde\rho}X(x)]A(f^nx)\eqno{(2)}  $$
Note that $\int dx\,\tilde\rho(x)=1$ implies $\int\rho(dx)\,(\tilde\rho'/\tilde\rho)(x)=\int dx\,\tilde\rho'(x)=0$.  Therefore, imposing mixing gives that
$$	\int dx\,\tilde\rho'(x)A(f^{N+1}x)
	=\int\rho(dx)\,(\tilde\rho'/\tilde\rho)(x)A(f^{N+1}x)  $$
tends to 0 when $N\to\infty$.  Equation $(2)$ now implies that
$$	\Psi(z)=-\sum_{n=0}^\infty z^n
	\int dx\,\tilde\rho(x)[{\rm div}_{\tilde\rho}X(x)]A(f^nx)  $$
is well defined for $z=1$, and $\int dx\,\tilde\rho'(x)A(x)=\Psi(1)$.
\medskip
	{Conclusion: }{\sl Suppose that $\rho$ is $f$-ergodic, with density $\tilde\rho$, and that $(f,\rho)$ is mixing on a function space ${\cal S}$ containing $\tilde\rho'/\tilde\rho$ and $A$, then $\Psi(1)$ is well defined, and $\delta_X\rho(A)=\int dx\,\tilde\rho'(x)A(x)=\Psi(1)$.}
\medskip\noindent
{\bf 2. Computer simulations.}
\medskip
	It is accepted that, using a computer, one can approximate numerically an SRB measure  by a time average:
$$	{1\over N_1-N_0}\sum_{n=N_0+1}^{N_1}\delta_{f^nx}      $$
for large $N_1-N_0$ (and $N_0$ moderately large); the idea is that the computed orbit $f^nx$ is noisy because of roundoff errors, and that this noisy orbit has an SRB time average\footnote{$^7$}{See [15] and, for example, Eckmann and Ruelle [6].}.  The Lyapunov exponents, and the coefficient $L_+$ introduced below, can also in principle be determined numerically.  It is therefore possible to estimate $r(\Psi)$ in particular cases, and to test the relations proposed above between the stable dimension $d$ of $\rho$ and the convergence radius $r(\Psi)$ in the presence of tangencies.  For example, let dim $M=2$, and let the Lyapunov exponents $\lambda_-,\lambda_+$ of $(f,\rho)$ satisfy $\lambda_-<0<\lambda_+$, so that\footnote{$^8$}{See L.-S. Young [24].} $d=\lambda_+/|\lambda_-|$.  Does the presence of tangencies together with $\lambda_+/|\lambda_-|<1/2$ imply $r(\Psi)<1$? (This appears to be the case for the classical H\'enon attractor).  Does $\lambda_+/|\lambda_-|\ge1/2$ imply $r(\Psi)\ge1$?  Are there examples with tangencies and $r(\Psi)>1$?
\medskip\noindent
{\bf 3. Singularities of $\Psi$ in the presence of tangencies.}
\medskip
	It is readily seen that the power series (1) defining the susceptibility has a radius of convergence $r(\Psi)>0$.  Tangencies between stable and unstable manifolds for $(f,\rho)$ are expected to produce singularities of $\Psi$, thus limiting $r(\Psi)$.  A difficulty of the problem is that the set of points of tangency has measure zero.  Note in this respect that the angle between stable and unstable manifolds is defined $\rho$-a.e., and that the a.e. range of this angle determines if tangencies are allowed or not.  A similar comment can be made for higher order contacts of the stable-unstable manifolds.  It is reasonable to  exclude those higher order contacts which (given the dimension of $M$) are nongeneric if the stable and unstable manifolds are regarded as independent.  At a generic tangency point $O$, the unstable manifold is folded in a way which is basically 2-dimensional (corresponding to variables $x,y$ introduced below).  Along the orbit $(f^nO)$ we have folds which are sharper and sharper as $n\to\infty$.  This exponential sharpening of the folds, combined with the derivative $\partial_x$ in $(1)$, produces the singularities of $\Psi(z)$ with $|z|<1$ which we want to study.
\medskip
	One can prove that $r(\Psi)<1$ for certain unimodal maps of the interval\footnote{$^{9}$}{See footnote 5.}; these maps are non-invertible and give a degenerate example of tangencies that is relatively accessible to mathematical study.  In what follows we discuss a crude imitation of the 1-dimensional situation for higher-dimensional diffeomorphisms.  In the case of unimodal maps of the interval with an ergodic measure $\rho$ absolutely continuous with respect to Lebesgue, the density of $\rho$ has {\it spikes} $\sim|x-f^nc|^{-1/2}$ on one side of the points $f^nc$ of the postcritical orbit.  These spikes are at the origin of the singularities of $\Psi(z)$ inside of the unit circle.  Instead of an individual postcritical point, we find for higher dimensional diffeomorphisms a family of tangencies of stable and unstable manifolds: think of a pile of (local) unstable manifolds (with tangencies) carrying part of the measure $\rho$.  Morally, this means that the spikes are ``spread out'' or ``smoothed'' (corresponding to integration over a measure transverse to the unstable manifolds).  This smoothing may give weaker singularities of $\Psi$ (i.e., larger $r(\Psi)$).
\medskip
	Let us choose local coordinates $(x,X,y,Y)\in{\bf R}\times{\bf R}^{s-1}\times{\bf R}\times{\bf R}^{u-1}$ such that the $s$-dimensional stable manifolds are given by $(y,Y)=$ const., and the local unstable manifold $U$ through $O$ is given by $x=ay^2,X=0,Y=0$ (for definiteness we take $a>0$).  The conditional measure of $\rho$ on $U$ is thus $\sim\Delta(dx\,dX\,dy\,dY)=\delta(x-ay^2)\delta(X)\delta(Y)dx\,dX\,dy\,dY$.  One can argue that the variable $Y$ does not play an important role in the present discussion, and we shall omit it, which amounts to taking $u=1$.  Using similar local coordinates near $fO$, we assume that the map $f$ has the form
$$	(x,X,y)\mapsto(e^{L_+}x,e^\Lambda X,e^{L_-}y)      $$
where $L_-<0,L_+>0$, and $e^\Lambda$ is a contraction (stronger than that given by $e^{L_-}$).
\medskip
	The assumption that the unstable manifolds are parallel affine manifolds is crude, and so is the assumption that $L_-,L_+$, and $\Lambda$ are constant coefficients.  [One might think of $L_-,L_+$, and $\Lambda$ as Lyapunov exponents.  But $L_+$, the only one of these coefficients to appear in the final formulas, is really the mean rate of expansion along a forward orbit $(f^nO)$ of tangencies].  These crude assumptions may be in part justified by the fact that we are looking for the leading singular behavior associated with a subset of unstable manifolds.  We shall use informally the notation $\approx$ (approximately equal to) and $\sim$ (approximately proportional to) in trying to find the leading singular behavior.
\medskip
	The contribution of the conditional measure $\Delta$ of $\rho$ on the piece $U$ of unstable manifold is
$$	\Psi^\Delta(z)\sim\sum_{n=0}^\infty z^n
	\int\Delta(dx\,dX\,dy)\,{\bf X}(x,X,y)\cdot\partial_{(x,X,y)}(A\circ f^n)   $$
Singularities for $|z|\le1$ can only come from the component ${\bf X}_1$ of ${\bf X}$ in the $x$-direction, giving
$$	\Psi^\Delta(z)\sim\sum_{n=0}^\infty z^n
	\int dx\,dy\,\delta(x-ay^2){\bf X}_1\partial_x(A_1\circ f^n)      $$
$$	\approx\sum_{n=0}^\infty(ze^{L_+})^n{\bf X}_1(0)
	\int (f^{*n}\delta(x-ay^2)dx\,dy)A'_1(x)      $$
where $A'_1(x)$ is the derivative of $A_1(x)$, which is $A$ evaluated in the coordinates $(x,0,0)$ centered at $f^nO$.
\medskip
	One has
$$	\delta(x-ay^2)={1\over2\sqrt{ax}}
	[\delta(y-{\sqrt{x}\over\sqrt{a}})+\delta(y+{\sqrt{x}\over\sqrt{a}})]  $$
hence
$$	f^{*n}(\delta(x-ay^2)dx\,dy)
	={e^{-nL_+/2}\over2\sqrt{ax}}[\delta(y-{\sqrt{x}\over\sqrt{a_n}})
	+\delta(y+{\sqrt{x}\over\sqrt{a_n}})]dx\,dy      $$
where $a_n=ae^{nL_+}e^{-2nL_-}$.  Therefore
$$	\Psi^\Delta(z)\approx
	\sum_{n=0}^\infty(ze^{L_+/2})^n
	{{\bf X}_1(0)\over\sqrt{a}}\int_0^{\rm cutoff}{dx\over\sqrt{x}}A'_1(x)    $$
so that $r(\Psi^\Delta)=e^{-L_+/2}<1$.  This result is in agreement with that obtained with the spikes of the invariant density for unimodal maps in 1 dimension (which are limiting cases of diffeomorphisms with tangencies).
\medskip
	Remember however that $\Psi$ is defined with the measure $\rho$ rather than $\Delta$.  Let thus $\Gamma$ be part of the measure $\rho$, carried by a pile of unstable manifolds (with tangencies) near $O$, and write
$$	\Gamma(dx\,dX\,dy)
	=\int\gamma(d\xi\,dX)\,\delta(x-a(\xi,X)(y-b(\xi,X))^2-c(\xi,X))dx\,dy$$
where the integration is over the variable $\xi$, and $\gamma(d\xi\,dX)$ is a transverse measure of $\rho$ in the stable direction, and we assume $a(\xi,X)>0$.  It will turn out that we obtain the same estimate of $r(\Psi^\Gamma)$ for different $\Gamma$'s, and we expect that the contributions $\Psi^\Gamma$ to $\Psi$ of different $\Gamma$'s will add up convergently for $|z|<r(\Psi^\Gamma)$.  [Such behavior was found for the contributions of different spikes in the unimodal case].  The most singular part of $\Psi^\Gamma$ is of the form
$$	\Psi_1^\Gamma(z)=\sum_{n=0}^\infty z^n
	\int\Gamma(dx\,dX\,dy){\bf X}_1(x,X,y)\partial_x(A_1\circ f^n)      $$
$$	\approx\sum_{n=0}^\infty z^n
	\int\gamma(d\xi\,dX)\,\delta(x-a(\xi,X)(y-b(\xi,X))^2-c(\xi,X))
	{\bf X}_1(x,X,y)\partial_xA_1(e^{nL_+}x)dx\,dy      $$
$$	=\sum_{n=0}^\infty(ze^{L_+})^n
	\int\gamma(d\xi\,dX)\,\delta(x-a(\xi,X)(y-b(\xi,X))^2-c(\xi,X))
	{\bf X}_1(x,X,y)A'_1(e^{nL_+}x)dx\,dy      $$
To define $A_1(x)=A(x,0,0)$ and $A_1'(x)$ we have again used coordinates $(x,0,0)$ centered at $f^nO$.  Note that $A_1'(e^{nL_+}x)=A_1'([f^n(x,0,0)]_1)$ where $[\cdot]_1$ denotes the first component.  Therefore, when $n$ is large, the functions $x\mapsto[f^n(x,0,0)]_1,A_1'(e^{nL_+}x)$ oscillate many times, with a frequency $\sim nL_+$.
\medskip	
	We may replace ${\bf X}_1(x,X,y)$ by $\tilde X(\xi,X)={\bf X}_1(c(\xi,X),X,b(\xi,X))$, and write
$$	\int\delta(x-a(\xi,X)(y-b(\xi,X))^2-c(\xi,X))dy
	={1\over\sqrt{a(\xi,X)}}\cdot{1\over\sqrt{x-c(\xi,X)}}  $$
where the right-hand side is replaced by 0 if $x<c(\xi,X)$.  Then
$$	\Psi_1^\Gamma(z)\approx\sum_{n=0}^\infty(ze^{L_+})^n
	\int{\gamma(d\xi\,dX)\tilde X(\xi,X)\over\sqrt{a(\xi,X)}\sqrt{x-c(\xi,X)}}
	A'_1(e^{nL_+}x)dx  $$
If we let $\tilde\gamma(d\tilde\xi)$ be the image of the measure $\gamma(d\xi\,dX)\tilde X(\xi,X)/\sqrt{a(\xi,X)}$ by $(\xi,X)\mapsto\tilde\xi=c(\xi,X)$, we obtain finally
$$	\Psi_1^\Gamma(z)\approx\sum_{n=0}^\infty(ze^{L_+})^n
	\int h(x)A'_1(e^{nL_+}x)dx\qquad{\rm where}\qquad 
	h(x)=\int{\tilde\gamma(d\tilde\xi)\over\sqrt{x-\tilde\xi}}  $$
\medskip\noindent
{\bf 4. Estimates when {\rm supp}$\tilde\gamma$ has zero Lebesgue measure.}
\medskip
	We assume now that supp$\tilde\gamma$ has Lebesgue measure $=0$.  Given $x\notin$ supp $\tilde\gamma$, let
$$	\xi^*={\rm max}\{\xi\in{\rm supp}\,\tilde\gamma:\xi<x\}.  $$
and let $\gamma^*(d\eta)$ be the image, restricted to $\eta\ge0$, of $\tilde\gamma(d\tilde\xi)$ by $\tilde\xi\mapsto\eta=\xi^*-\tilde\xi$.  We have then
$$	h(x)=\int{\gamma^*(d\eta)\over\sqrt{(x-\xi^*)+\eta}}
	=\int{\phi'(\eta)\,d\eta\over\sqrt{(x-\xi^*)+\eta}}  $$
where $\phi(\eta)=\int_0^\eta\gamma^*(dt)$ and $\phi(\eta)\sim\eta^d$ for small $\eta$.  If $0<\alpha<1-d$ we let
$$	h_1(x)=\int((x-\xi^*)+\eta)^{\alpha-1}\phi'(\eta)d\eta
	=(1-\alpha)\int_0^\infty((x-\xi^*)+\eta)^{\alpha-2}\phi(\eta)d\eta  $$
where we have put an upper limit $+\infty$ to the integral because it does not need a cutoff.  Therefore
$$	h_1(x)\approx(1-\alpha){(x-\xi^*)^{1+d}\over(x-\xi^*)^{2-\alpha}}
	\int_0^\infty{\phi(t)\,dt\over(1+t)^{1+\alpha}}=C_\alpha(x-\xi^*)^{d+\alpha-1}   $$
\indent
	{\bf A.} Assuming $d<1/2$ and taking $\alpha=1/2$ we may thus conclude that $h(x)\approx C(x-\xi^*)^{d-1/2}$, hence, if $I=(\xi^*,\xi^{**})$ is an interval of length $\ell$ of the complement of supp$\gamma$, we may estimate
$$ \int_I|h(x)|^pdx\approx C^p\int_0^\ell dt\,t^{-p(1/2-d)}=C'\ell^{1-p(1/2-d)}  $$
if $1\le p<(1/2-d)^{-1}$.  By scaling we assume that the number of intervals $I$ with $|I|\approx\ell$ is $\sim\ell^{-d}$.  We have thus
$$	\int|h(x)|^pdx\sim\sum_\ell\ell^{1-d-p(1/2-d)}  $$
If $p\ge1$, and $1-d-p(1/2-d)>0$, we have thus $h\in L_p$ (and the bound $p<(1-d)/(1/2-d)$ appears best possible).  Let $1/p+1/q=1$, then the Fourier transform $\hat h$ is in  $L_q$.  We have
$$	{1\over q}<1-{1/2-d\over1-d}={1/2\over1-d}\eqno{(3)}  $$
(and this bound appears best possible).  If $|\hat h(s)|\sim s^{-t}$ for large $s$, we need $tq>1$, i.e., $t>1/q$ if $1/q$ satisfies $(3)$, i.e.,
$$	|\hat h(s)|\sim s^{-t}\qquad{\rm with}\qquad t\ge{1/2\over1-d}  $$
We come now to the estimation of $\int h(x)A_1'(e^{nL_+}x)\,dx$ where, for large $n$, $x\mapsto A_1'(e^{nL_+}x)$ is rapidly oscillating with frequency $\sim nL_+$.  Since we are interested in the most singular part of $h$, we may replace it by a function with compact support.  Because $A'_1$ is a derivative, there is no zero-frequency contribution to the integral, and we have
$$	\int h(x)A_1'(e^{nL_+}x)\,dx\sim\hat h(e^{nL_+})  $$
with a negligible contribution of higher harmonics.  Therefore
$$	|\int h(x)A'_1(e^{nL_+}x)\,dx|\sim|\hat h(e^{nL_+})|
	\sim e^{-tnL_+}\le\exp[-n{1/2\over1-d}L_+]  $$
and the bound again appears best possible, so that $\Psi_1^\Gamma(z)$ converges for
$$	|z|<\exp[-(1-{1/2\over1-d})L_+]
	=\exp[-{1/2-d\over1-d}L_+]  $$
i.e., $r(\Psi_1^\Gamma)=\exp[-(1/2-d)L_+/(1-d)]$, and a reasonable guess would appear to be 
$$  r(\Psi)=\exp[-{1/2-d\over1-d}L_+]   $$
[hence $e^{-L_+/2}<r(\Psi)<\exp[(d-1/2)L_+]<1$].
\medskip\indent
	{\bf B.} Assuming $1/2\le d<1$, we write $h$ (which is the convolution product $\tilde\gamma*(\cdot)^{-1/2}$) as 
$$	h\sim\tilde\gamma*(\cdot)^{\alpha-1}*(\cdot)^{\beta-1}  $$
where $\alpha,\beta>0,\alpha+\beta=1/2,d+\alpha<1$, or $\beta=1/2-\alpha,0<\alpha<1-d$.  We have thus
$$	h=h_1*(\cdot)^{\beta-1}\qquad{\rm where}
	\qquad h_1=\tilde\gamma*(\cdot)^{\alpha-1}  $$
and we have seen that $h_1(x)\approx C_\alpha(x-\xi^*)^{d+\alpha-1}$.  We find as in {\bf A.} that we can take $|\hat h_1(s)|\sim s^{-t}$ with $t\ge\alpha/(1-d)$, hence
$$	\hat h(s)\sim s^{-\alpha/(1-d)}s^{-\beta}
	=s^{-1/2-\alpha d/(1-d)}  $$
so that 
$$	|\int h(x)A_1'(e^{nL_+}x)\,dx|\sim|\hat h(e^{nL_+})|
	\le\exp[-n({\alpha d\over1-d}+{1\over2})L_+]  $$
and $\Psi_1^\Gamma(z)$ converges for
$$	|z|<\exp[({\alpha d\over1-d}+{1\over2}-1)L_+]
	=\exp[({\alpha d\over1-d}-{1\over2})L_+]  $$
where we may let $\alpha\to1-d$, hence we may estimate
$$	r(\Psi_1^\Gamma)\ge\exp[({\alpha d\over1-d}-{1\over2})L_+]
	=\exp[(d-1/2)L_+]  $$
which is $>1$.  In fact $r(\Psi_1^\Gamma)=\exp[(d-1/2)L_+]$ is a reasonable guess.
\medskip
	The convergence radius $r(\Psi)$ now depends on the behavior of $(f,\rho)$ away from tangencies, and we may expect that the derivative $\partial_x$ in $(1)$ plays a less important role.  Therefore $r(\Psi)$ is expected to depend on the mixing properties of $(f,\rho)$, over which one has some control [25], [26].  One may thus hope that $r(\Psi)\ge1$, or even $r(\Psi)>1$, and that $\Psi(1)$ is well defined.  The situation where the set of tangencies is large ($d>1/2$) reminds one of Newhouse's study of persistent tangencies (wild hyperbolic sets, infinitely many sinks, see [11], [12], [13]).  While the situation considered by Newhouse has very discontinuous topology, it is not unthinkable that the particular measure $\rho$ behaves differentiably in some sense.
\medskip\indent
	{\bf C.} It is plausible that the results of {\bf A.} and {\bf B.} remain true without the condition that supp$\tilde\gamma$ has zero Lebesgue measure.  Furthermore, if the stable dimension $d$ of $\rho$ is $\ge1$, one can write $d$ as a sum of partial dimensions\footnote{$^{10}$}{See [6] Section IV.D, and references given there, in particular [10].}, and use arguments as above.  One expects thus that the formula $r(\Psi_1^\Gamma)\ge\exp[(d-1/2)L_+]$ will remain correct in that case and, as argued in {\bf B.}, we may then have $r(\Psi)\ge1$ or even $r(\Psi)>1$.
\medskip
	If we have a continuous time dynamical system (a flow) instead of discrete time dynamics (a diffeomorphism), we expect similar results in the presence of tangencies: a susceptibility function $\hat\kappa(\omega)$ with singularities in the upper half $\omega$-plane if $d<1/2$, no singularity if $d\ge1/2$, and $\hat\kappa(0)$ hopefully well defined if $d>1/2$.  The continuous time dynamical situation is that most relevant for physical applications.
\medskip\noindent
{\bf 5. Physical discussion.}
\medskip
	In this brief physically oriented discussion we shall, for simplicity, use the language of discrete time dynamical systems.
\medskip
	We have made above a nonrigorous analysis of how tangencies between stable and unstable manifolds may influence the radius of convergence $r(\Psi)$ of the susceptibility function.  We have found two different regimes depending on whether the stable dimension $d$ of the SRB measure $\rho$ is $<1/2$ or $\ge1/2$.
\medskip
	If $d<1/2$ we expect $r(\Psi)<1$, i.e., the tangencies cause singularities of $\Psi(z)$ with $|z|<1$.  Such singularities reflect the exponential growth of small periodic perturbations of the dynamics $(f,\rho)$ (see [20]).  Experimentally, this may be visible as resonant behavior when a physical system is excited by a weak periodic signal: it would be of particular interest to study the case of hydrodynamic turbulence.
\medskip
	If $d\ge1/2$ we expect $r(\Psi)\ge1$ and, if $d>1/2$, the value $\Psi(1)$ may be well defined.  Since $\Psi(1)$ is formally related to the derivative of $\rho$ with respect to $f$, we may hope that this derivative exists in some sense.  This would apply to physical systems not too far from equilibrium (at equilibrium, $\rho$ has a density, and $d\ge1$ unless all Lyapunov exponents vanish) with obvious application to linear response in nonequilibrium statistical mechanics.  For large physical systems (dim$M$ large), when there is chaos and a density of Lyapunov exponents can be defined, one also expects $d$ large by the Kaplan-Yorke formula\footnote{$^{11}$}{See [6] Section IV.C, and references given there, in particular [7].}, provided the degrees of freedom of the large system have a sufficiently strong effective interaction.
\medskip
      In view of the mathematical difficulty of analyzing dynamical systems with tangencies, a computer-experimental study would be desirable.  The situation of choice would be that of 2-dimensional diffeomorphisms with an SRB measure $\rho$ such that the Lyapunov exponents $\lambda_-,\lambda_+$ satisfy $\lambda_-<0<\lambda_+$.  In that case we know [24] that $d=\lambda_+/|\lambda_-|$, and the radius of convergence $r(\Psi)$ is also accessible numerically.  For the classical H\'enon attractor we have $d<1/2$, and it appears [4] that $r(\Psi)<1$.  In other cases, studied by Ueda and coworkers [22], [23], visual inspection of the computer plot of the attractor seems to indicate a large $d$, and it would be desirable to estimate $r(\Psi)$.
\medskip\noindent
{\bf 6. Infinitesimally stable ergodic measures.}
\medskip
	Consider the general situation of a diffeomorphism $f$ of the compact manifold $M$, and of an ergodic measure $\rho$ for $f$ on $M$.  We want to study formally the stability of $\rho$ under an infinitesimal change of $f$.
\medskip
	We shall use a space ${\cal D}$ of smooth functions on $M$, with dual ${\cal D}^*$, and a space ${\cal V}$ of smooth vector fields on $M$.  If $X\in{\cal V}$, we write $\hat X(A)=\int\rho(dx)\,X(x)\cdot\partial_xA$, so that $\hat X\in{\cal D}^*$.  Defining $T:{\cal D}^*\to{\cal D}^*$ and $Tf:{\cal V}\to{\cal V}$ by
$$	(T\xi)(A)=\xi(A\circ f)\qquad,\qquad((Tf)X)(fx)=(T_xf)X(x)   $$
we find that $((Tf)X)^\wedge=T\hat X$.
\medskip
	Consider $\rho+\hat X$ as an infinitesimal perturbation of $\rho$ (it corresponds to replacing $\rho$ by its image under $x\mapsto x+X(x)$).  The measure $\rho$ is mapped to itself by $f$, while $\rho+\hat X$ is mapped to $\rho+T\hat X$.  We say that $\rho$ is {\it infinitesimally stable} (or attracting) if $(T^n\hat X)(A)\to0$ exponentially\footnote{$^{12}$}{An alternate (weaker) requirement would be that $\Psi(1)=\sum_{n=0}^\infty\rho(dx)\,X(x)\cdot\partial_x(A\circ f^n)$ converges whenever $X\in{\cal V},A\in{\cal D}$.} with $n$ whenever $X\in{\cal V},A\in{\cal D}$.  It is plausible that an infinitesimally stable measure must be SRB.
\medskip
	We perturb $f$ to $\tilde f=f+X\circ f$, where $X\in{\cal V}$.  If $\xi\in{\cal D}^*$, the $\tilde f$-invariance of $\rho+\xi$, i.e.,
$$	(\rho+\xi)(A\circ(f+X\circ f))=(\rho+\xi)(A)   $$
is then given, to first order in $X$, by
$$	\int\rho(dx)\,[A(fx)+X(fx)\cdot\partial_{fx}A]+\xi(A\circ f)=\rho(A)+\xi(A)   $$
or $\hat X+T\xi=\xi$, hence $T^n\xi-T^{n+1}\xi=T^n\hat X$, hence $\xi-T^{N+1}\xi=\sum_{n=0}^NT^n\hat X$.  Therefore, if $\rho$ is infinitesimally stable, we obtain $\rho+\xi$ which is $\tilde f$-invariant to first order by taking
$$	\xi(A)=\sum_{n=0}^\infty(T^n\hat X)(A)   $$
and $\xi$ is unique such that $(T^n\xi)(A)\to0$ for all $A\in{\cal D}$ when $n\to\infty$.  This shows that the linear response $X\mapsto\xi$ is related to infinitesimal stability.
\medskip
	The above considerations apply to the uniformly hyperbolic situation where $\rho$ is an SRB measure on an Axiom A attractor.  The purpose of the present paper has been to make plausible the infinitesimal stability of SRB measures in a different situation where there are stable-unstable tangencies.  [Note that $\Psi(z)=\sum_{n=0}^\infty z^n(T^n\hat X)(A)$, so that $r(\Psi)>1$ is equivalent to the infinitesimal stability condition that $(T^n\hat X)(A)\to0$ exponentially].
\medskip\noindent
{\bf References.}
\medskip
[1] A. \`Avila, S. Gou\"ezel, and M. Tsujii  ``Smoothness of solenoidal attractors.'' Discr. and Cont. Dynam. Syst. {\bf 15},21-35(2006).

[2] C. Bonatti, L. Diaz, and M. Viana.  {\it Dynamics beyond uniform hyperbolicity.} Sprin\-ger, Berlin, 2005.

[3] O. Butterley and C. Liverani  ``Smooth Anosov flows: correlation spectra and stability.''  J. Modern Dynamics {\bf 1},301-322(2007).

[4] B. Cessac  ``Does the complex susceptibility of the H\'enon map have a pole in the upper half plane?  A numerical investigation.''  Nonlinearity {\bf 20},2883-2895(2007).

[5] D. Dolgopyat  ``On differentiability of SRB states for partially hyperbolic systems.''  Invent. Math. {\bf 155},389-449(2004).

[6] J.-P. Eckmann and D. Ruelle  ``Ergodic theory of chaos and strange attractors.''  Rev. Mod. Phys. {\bf 57},617-656(1985).
\goodbreak

[7] P. Frederickson, J.L. Kaplan, E.D. Yorke, and J.A. Yorke  ``The Lyapunov dimension of strange attractors.''  J. Diff. {\bf 49},185-207(1983).

[8] Y. Jiang and D. Ruelle  ``Analyticity of the susceptibility function for unimodal Markovian maps of the interval.''  Nonlinearity {\bf 18},2447-2453(2005).

[9] A. Katok, G. Knieper, M. Pollicott, and H. Weiss  ``Differentiability and analyticity of topological entropy for Anosov and geodesic flows.''  Invent. Math. {\bf 98},581-597(1989).

[10] F.Ledrappier and L.S.Young.  ``The metric entropy of diffeomorphisms: I. Characterization of measures satisfying Pesin's formula, II. Relations between entropy, exponents and dimension.''  Ann. of Math. {\bf 122},509-539,540-574(1985).
  
[11] S. Newhouse  ``Diffeomorphisms with infinitely many sinks.'' Topology  {\bf 13},9-18 (1974).

[12] S. Newhouse  ``The abundance of wild hyperbolic sets and nonsmooth stable sets for diffeomorphisms.''  Publ. Math. IHES {\bf 50},102-151(1979).

[13] S. Newhouse  ``Lectures on dynamical systems." pp. 1-114 in: {\it CIME Lectures, Bressanone, Italy, June 1978.}  Birkh\"auser, Boston, 1980.

[14] J. Palis and J.-P. Yoccoz  ``Non-uniformly hyperbolic horseshoes arising from bifurcations of heteroclinic cycles.''  IHES Publications Math\'ematiques {\bf 110},1-217(2010).

[15] D. Ruelle   ``What are the measures describing turbulence?''  Progr. Theoret. Phys. Suppl. {\bf 64},339-345(1978).

[16] D. Ruelle  ``Differentiation of SRB states.''  Commun. Math. Phys. {\bf 187},227-241(1997); ``Correction and complements.''  Commun. Math. Phys. {\bf 234},185-190(2003).

[17] D. Ruelle  ``Differentiating the absolutely continuous invariant measure of an interval map f with respect to f.''  Commun.Math. Phys. {\bf 258},445-453(2005).

[18] D. Ruelle  ``Differentiation of SRB states for hyperbolic flows.''  Ergod. Theor. Dynam. Syst. {\bf 28},613-631(2008).

[19] D. Ruelle  ``Structure and $f$-dependence of the a.c.i.m. for a unimodal map $f$ of Misiurewicz type.''  Commun. Math. Phys. {\bf 287},1039-1070(2009).

[20] D. Ruelle  ``A review of linear response theory for general differentiable dynamical systems.''  Nonlinearity {\bf 22},855-870(2009).

[21] J. Toll  ``Causality and the dispersion relation: logical foundations.''  Phys. Rev. {\bf 104},1760-1770(1956).

[22] Y. Ueda  ``Explosion of strange attractors exhibited by Duffing's equation.''  Ann. NY Acad. Sci. {\bf 357},422-434(1980).

[23] Y. Ueda and N. Akamatsu  ``Chaotically transitional phenomena in the forced negative resistance oscillator.''  IEEE Trans. CAS-{\bf 28},217-223(1981).

[24] L.-S. Young  ``Dimension, entropy and Liapunov exponents.''  Ergod. Theory Dynam. Syst. {\bf 2},109-124(1982).

[25] L.-S. Young  ``Statistical properties of dynamical systems with some hyperbolicity.'' Annals of Math. {\bf 147},585-650(1998).

[26] L.-S. Young  ``Recurrence times and rates of mixing.'' Israel J. Math. {\bf 110},153-188(1999).

[27] L.-S. Young  ``What are SRB measures, and which dynamical systems have them?''  J. Statist. Phys. {\bf 108},733-754(2002).

\end